\documentclass[aps,prb,preprint,superscriptaddress]{revtex4-1}
\usepackage{epsfig}
\usepackage{graphicx}
\usepackage[caption=false]{subfig}
\usepackage{bm}
\usepackage{color}

\begin{document}


\title{Toward the Mott state with Magnetic Cluster Formation in Heavily Cu-Doped NaFe$_{1-x}$Cu$_{x}$As}


\author{Yizhou Xin}
\affiliation{Department of Physics and Astronomy, Northwestern University, Evanston IL 60208, USA}
\author{Ingrid Stolt}
\affiliation{Department of Physics and Astronomy, Northwestern University, Evanston IL 60208, USA}
\author{Jeongseop A. Lee}
\affiliation{Department of Physics and Astronomy, Northwestern University, Evanston IL 60208, USA}
\author{Yu Song}
\altaffiliation{Present address: Department of Physics, University of California, Berkeley CA 94720, USA }
\affiliation{Department of Physics and Astronomy and Rice Center for Quantum Materials, Rice University, Houston TX 77005, USA}

\author{Pengcheng Dai}
\affiliation{Department of Physics and Astronomy and Rice Center for Quantum Materials, Rice University, Houston TX 77005, USA}
\author{W. P. Halperin}
\affiliation{Department of Physics and Astronomy, Northwestern University, Evanston IL 60208, USA}

\date{\today}

\begin{abstract}
Recent neutron scattering measurements indicate that NaFe$_{1-x}$Cu$_{x}$As forms an antiferromagnetic stripe phase near $x\approx 0.5$ in a Mott insulating state.  This copper concentration is well in excess of that required for superconductivity, $x < 0.04$.  We have investigated  the development of magnetism in this compound using $^{23}$Na nuclear magnetic resonance (NMR) spectra and spin-lattice relaxation measurements performed on single crystals ($x$ = 0.13, 0.18, 0.24, and 0.39). We find multiple inequivalent Na sites, each of which is associated with a different number of nearest neighbor Fe sites occupied by a Cu dopant. We show that the distribution of Cu substituted for Fe is  random in-plane for low concentrations ($x = 0.13$ and 0.18), but deviates from this with increasing Cu doping.  As is characteristic of many pnictide compounds, there is a spin pseudo gap that increases in magnitude with dopant concentration.  This is correlated with a corresponding increase in orbital NMR frequency shift indicating a  change in valence from Cu$^{2+}$ to a Cu$^{1+}$ state as $x$ exceeds 0.18, concomitant with the change of Fe$^{2+}$ to Fe$^{3+}$ resulting in the formation of magnetic clusters.  However, for $x\leq 0.39$ there is no evidence of long-range static magnetic order.  

\end{abstract}


\maketitle


\section{Introduction}
\begin{figure}[t]
	\includegraphics[scale = 0.5]{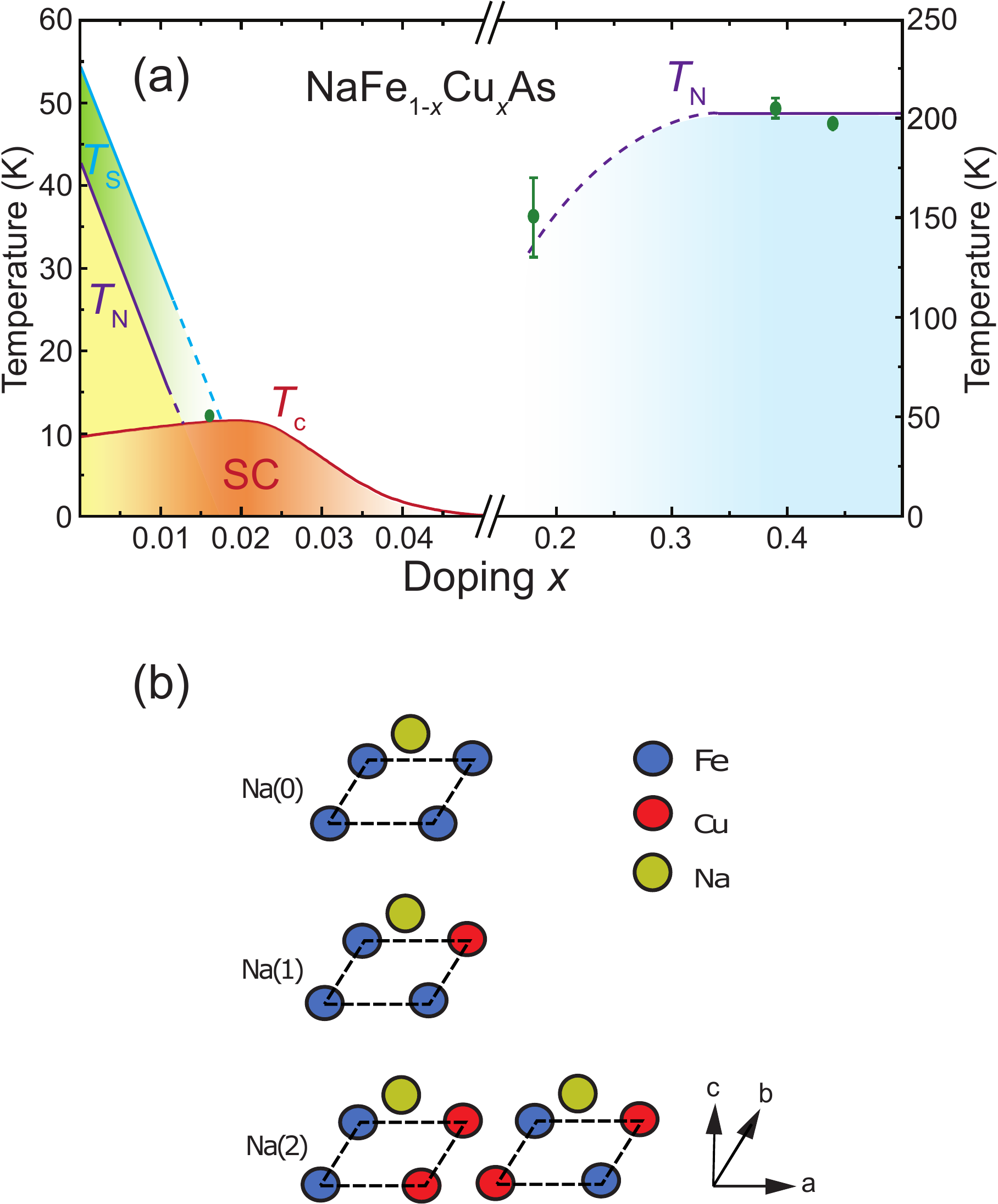}
	\caption{\label{Fig.1} (a) Phase Diagram of  NaFe$_{1-x}$Cu$_{x}$As single crystals showing superconducting regions at low copper doping and magnetic ordering at high doping levels; adapted from Ref.~\citenum{Wan.13} and ~\citenum{Son.15}. For $x \geq 0.18 $, $T_N$ is approximated from the onset temperature of normalized intensity from neutron scattering measurements.~\cite{Son.15,Son.16} (b) Schematic of Na sites with various environments according to the number of nearest neighbor Cu dopants substituted for Fe. The Na nuclei are located either above or below the Fe layer.}
\end{figure}

A major task in understanding superconductivity in Fe-based superconductors is to investigate its relationship with antiferromagnetism (AFM).~\cite{Dai.15,Ye.15} Unlike cuprate superconductors, whose parent compounds are Mott insulating, the iron-based superconductors are found to have a metallic parent state.~\cite{Ste.11,Joh.10} This provides a  different perspective for understanding the mechanism of high temperature superconductivity.~\cite{Ye.15} Song {\it et al.}~\cite{Son.16} recently reported that the heavily Cu-doped pnictide,  NaFe$_{1-x}$Cu${_x}$As (phase diagram  displayed in Fig.~\ref{Fig.1}\,(a)) exhibits real space Fe-Cu ordering with a Mott insulating antiferromagnetic stripe phase as $x$ approaches 0.5.  Their work was based on  transport, neutron scattering, transmission electron microscopy (TEM), X-ray absorption spectroscopy (XAS), and resonant inelastic X-ray scattering (RIXS) results. The Mott-insulating nature of heavily Cu-doped NaFe$_{1-x}$Cu${_x}$As is also confirmed by angle-resolved photoemission spectroscopy measurements.~\cite{Mat.16} It appears that the Cu-doped pnictides~\cite{Wan.13,Son.16,Wan.17} are a new class of compounds that exemplify important connections between antiferromagnetism (AFM) and superconductivity. 
 
Here we explore the  effects of Cu doping in high-quality NaFe$_{1-x}$Cu$_{x}$As (111) single crystals using nuclear magnetic resonance (NMR) as a microscopic probe of the local magnetic and electronic environment at the site of interest. In this paper, we report our  $^{23}$Na NMR investigations in the Cu doping  range,  $0.13 \leq x \leq$\,\,0.39. The $^{23}$Na and $^{75}$As nuclei are located on opposite sides of the Fe layer (Fig.~\ref{Fig.1}\,(b)) and provide complementary information about the role of the Cu dopant.  We focus on $^{23}$Na  which has a significantly narrower NMR linewidth than $^{75}$As due to its weaker hyperfine coupling to  electronic states that allows higher spectral resolution in probing distributions of local magnetic fields.  We show how the local magnetic and electronic properties of NaFe$_{1-x}$Cu$_{x}$As evolve with increasing Cu concentration as the system becomes more insulating and relate our NMR data to results from other experimental techniques, notably neutron scattering and magnetic susceptibility. This study will also provide the foundation for future work to better understand the nature of magnetic ordering in NaFe$_{1-x}$Cu$_{x}$As with $x\approx 0.5$.  

%
\section{Experimental Methods}
 NaFe$_{1-x}$Cu$_{x}$As single crystals were grown by a self-flux method at Rice University.\cite{Son.16} We performed $^{23}$Na NMR experiments for $x$ = 0.13, 0.18, 0.24, and 0.39  at Northwestern University and at the National High Magnetic Field Laboratory (NHMFL) in Tallahassee, Florida. Typically, for each doping we cleaved a $\sim$ 5$\times$5$\times$2 mm$^{3}$ piece from a larger single crystal  in a glove box filled with $^4$He gas of ultra-high purity.  Then the sample was placed in a leak tight sample holder machined from Stycast-1266 epoxy, filled with the same gas to avoid sample degradation from contact with air.~\cite{Oh.13b} The sample holder was designed to maximize the filling factor in the NMR coil. The majority of the data was collected with an Oxford 14 T superconducting magnet at Northwestern University, with the MagRes2000  wide-band spectrometer system designed by A. P. Reyes at the NHMFL. Each crystal was oriented with the $c$-axis parallel to the external magnetic field, $H_0$ = 13.98 T. For NMR measurements at low temperatures, we used a home-built continuous flow cryostat designed by J.A. Lee at Northwestern University, and an immersion style cryostat at NHMFL. The quadrupolar echo ($\pi$/2-$\pi$/2) NMR pulse sequence was used to obtain $^{23}$Na spectra of both central transition (1/2 $\leftrightarrow$ -1/2) and quadrupolar satellites (3/2 $\leftrightarrow$ 1/2 and -1/2 $\leftrightarrow$ -3/2), while the saturation recovery pulse sequence was used to measure the spin-lattice relaxation rates, 1/T$_1$. For the central transition, the $\pi/2$ pulse length was approximately 6 $\mu$s. By fitting gaussian models to the line shapes of the $^{23}$Na spectra, we extracted the Knight shift, linewidth and relative spin count.

There is an advantage to using NMR with a quadrupolar nucleus, such as $^{23}$Na (S = 3/2).  It has sensitivity to both charge and spin degrees of freedom. If the local environment at the nucleus has a non-zero electric field gradient (EFG),  usually from chemical bonding, then 2S - 1 satellites of the main spectral component called the central transition, will appear.  They are separated by a frequency splitting proportional to the product of the  EFG and the quadrupole moment of the nucleus that can be calculated in first order perturbation theory in the high magnetic field limit.  The maximum splitting of any orientation is the quadrupole frequency, $\nu_q$. The central transition, however, is only affected by the EFG in second order and can be neglected in our case. For the present work at high field this means that the  local magnetic field at the $^{23}$Na atom shifts all components of the three-fold spectrum identically and that the distribution in local magnetic fields contributes equally to the line widths of all three components.  However, in contrast, non-uniformity of the local EFG will significantly broaden the two quadrupolar satellites leaving the central transition largely unaffected. Even for high quality single crystals the satellites are usually broader than the central transition owing to the presence of dopants that give rise to a distribution in local EFG's,  which might make them unobservable, as shown by the faint intensity of the As quadrupolar satellites in the related compound Ba(Fe$_{0.93}$Co$_{0.07}$)$_2$As$_2$.~\cite{Oh.11} This is often the case for $^{75}$As NMR performed on pnictide compounds. For the Na atom in the 111 system, its quadrupolar satellites are usually better defined than that of As due to its smaller quadrupolar moment and weaker coupling to the dopant-induced disorder in the EFG.~\cite{Sto.98} Its hyperfine field is weak compared to that of $^{75}$As, based on our NMR results: $ \frac{ ^{75}A }{ ^{23}A} \approx 12$, consistent with the findings in the NMR study on NaFe$_{0.975}$Co$_{0.025}$As.~\cite{Oh.13b} Consequently using $^{23}$Na as the NMR spectator nucleus allows us to obtain independent information about the local chemical environment and magnetic order as  doping is increased.
 \section{NMR spectra}
\begin{figure}
	
	\subfloat{\includegraphics[scale = 0.32]{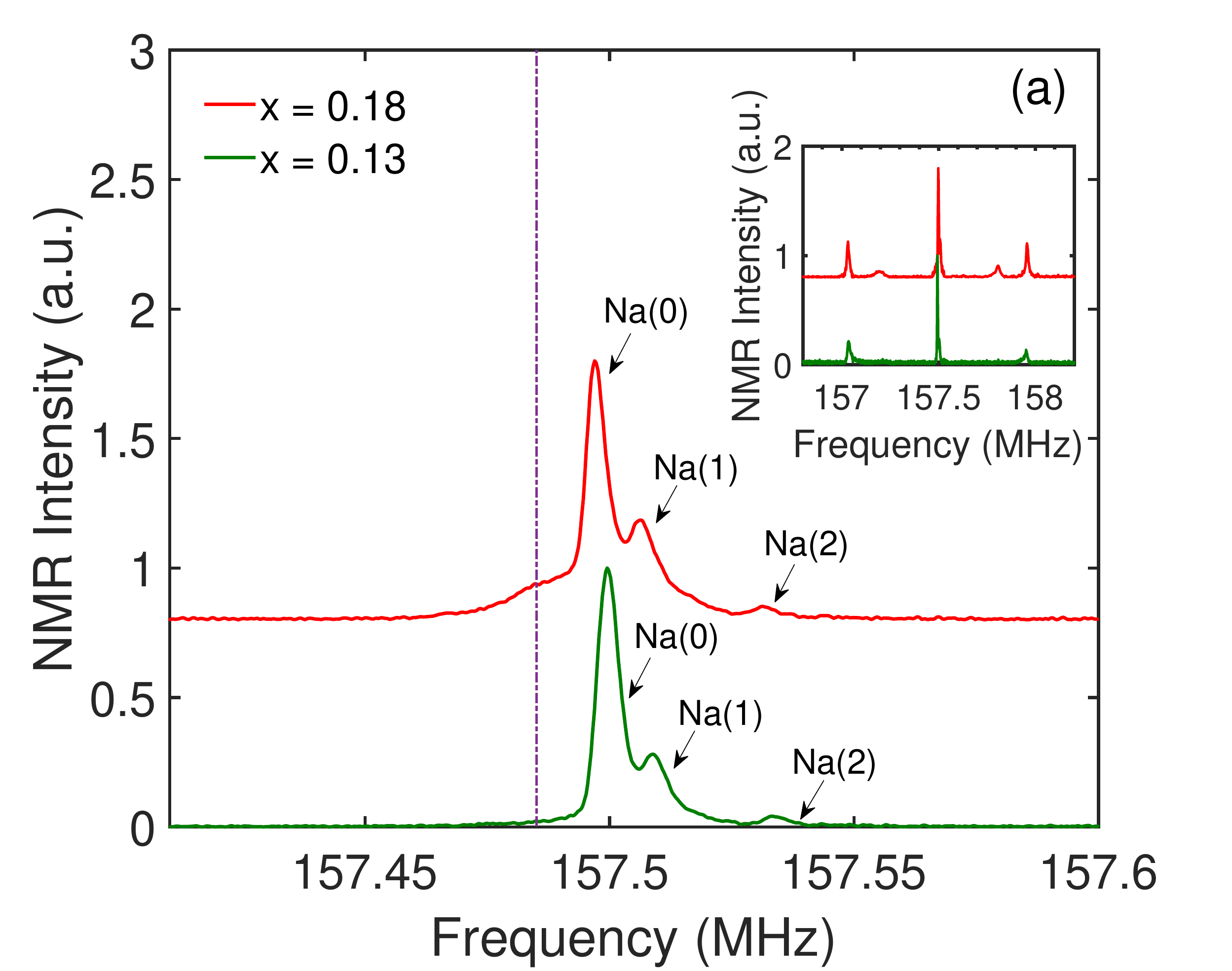}}
	\subfloat{\includegraphics[scale = 0.32]{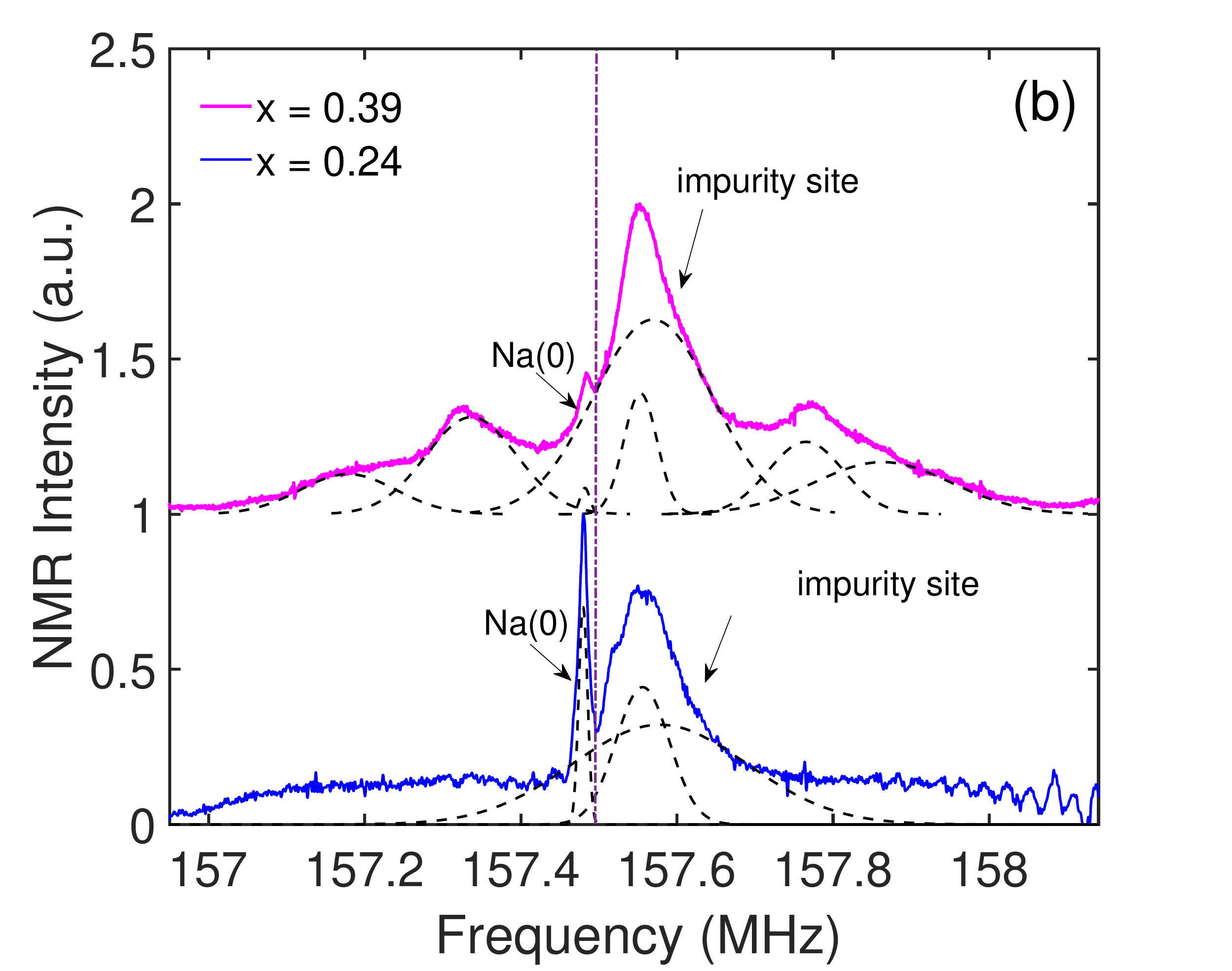}}
	\caption{\label{Fig.2} Room temperature $^{23}$Na NMR spectra of NaFe$_{1-x}$Cu$_{x}$As with four different Cu concentrations $x$ and external magnetic field $H_0= 14$\, T $\parallel$ $c$-axis: (a) $x=0.13$ and 0.18 and (b) $x=0.24$ and 0.39.  Three inequivalent $^{23}$Na sites, denoted by Na(0), Na(1), and Na(2), which are associated respectively with zero, one, and two NN Fe sites occupied by a Cu dopant, are found for the central transition (1/2 $\leftrightarrow$ -1/2). The full spectra, including the quadrupolar satellites, are shown in the inset, where $\nu_q$ is found to be 0.46 MHz for the Na(0) site and 0.31 MHz for the Na(1) site. Frequency sweeps were used to measure the full $^{23}$Na spectra of $x=0.24$ and 0.39. The fits to the spectra shown by dashed curves are described in the text. The purple vertical lines represent the Larmor frequency associated with $^{23}$Na.}
\end{figure}

\begin{figure}
	\includegraphics[scale =0.42]{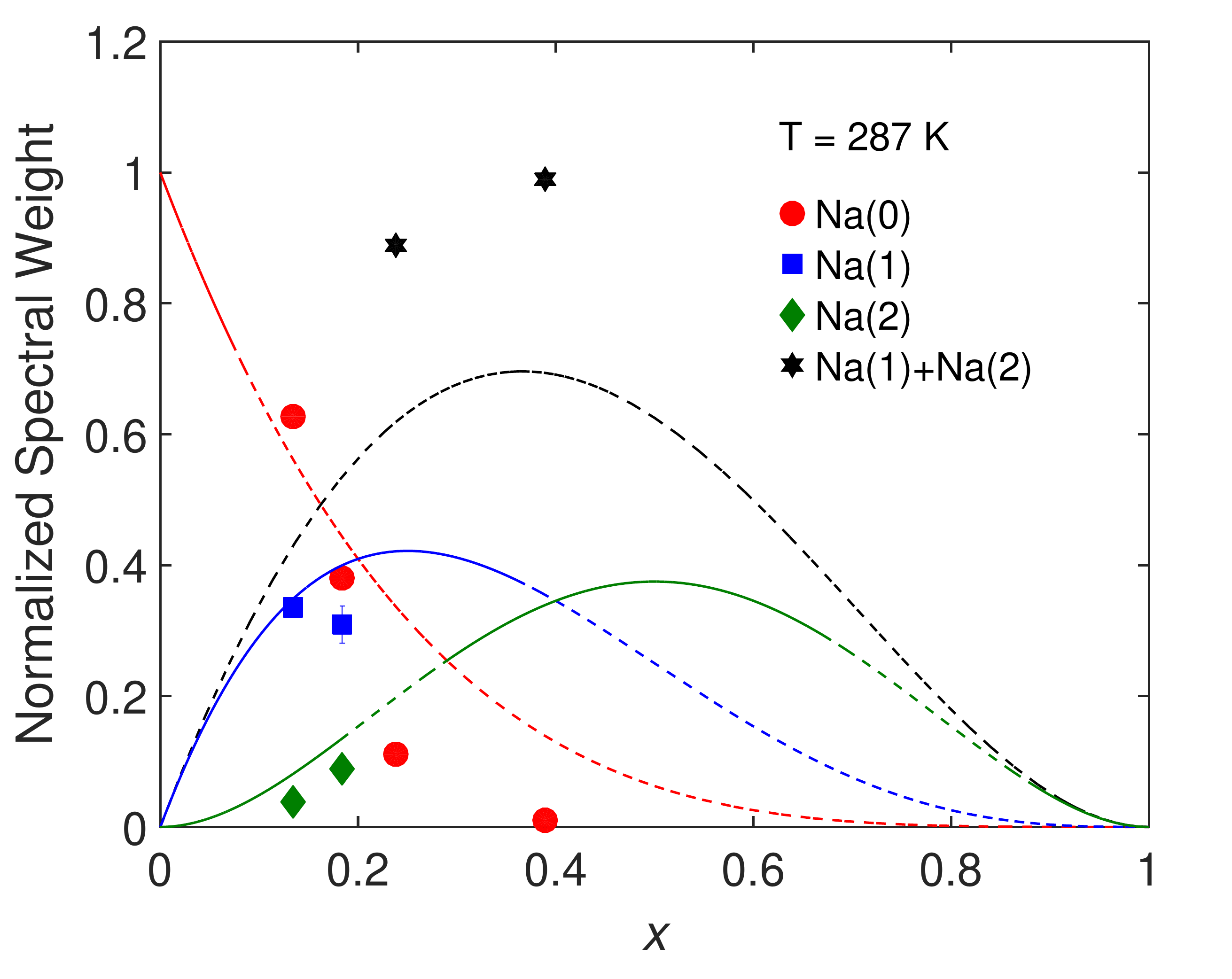}
	\caption{\label{Fig.3} Comparison between the experimental spectral weight of different $^{23}$Na sites and the spectral weight expected from the binomial model. The spectral weight is normalized to the total area under all components of the spectra. The normalized spectral weight of Na(0), Na(1), and Na(2) sites fit relatively well to the expected values from the binomial model for $x=0.13$  and 0.18; but for 0.24 and 0.39, the experimental values associated with the impurity site, which includes both Na(1) and Na(2) sites, far exceed the expected values from the binomial model, indicating that the Cu dopants are spatially correlated, a precursor to stripe formation.}
\end{figure}

Our room temperature $^{23}$Na  spectra are shown in Fig.~\ref{Fig.2}\,(a) for single crystalline samples $x=0.13$, and 0.18, and in Fig.~\ref{Fig.2}\,(b) for samples $x=0.24$ and 0.39. The dashed lines represent the Larmor angular frequency calculated from $\omega_{L}=\gamma H_0$, where the gyromagnetic ratio of $^{23}$Na is $\gamma /2\pi=11.2625$\,MHz/T.  We first discuss results for $x=0.13$ and 0.18 crystals.  As sketched in Fig.~\ref{Fig.1}\,(a) there are three distinct Na sites denoted by Na(0), Na(1), and Na(2).  They arise from Na sites having, respectively, zero, one, and two out of the four nearest neighbor (NN) Fe sites occupied by a Cu dopant.  This identification comes  from interpretation of the NMR spectral weight of the different components in the spectra in Fig.~\ref{Fig.2}\,(a) (spin count) that we discuss later in connection with Fig.~\ref{Fig.3}. 

The shifts of the central transition from the Larmor frequency are referred to as the Knight shift, a combination of spin shift and orbital shifts.  The spectra in the inset show  quadrupolar satellites symmetrically disposed relative to the central transition.  Similar observations of inequivalent Na sites have been reported from NMR measurements  on other systems, $e.g.$ in the 122-family.\cite{Nin.10,Nin.08} From the quadrupolar satellites for $x=0.13$ and 0.18 shown in the inset of Fig.~\ref{Fig.1}(a), we find the quadrupolar frequency splitting $\nu_q \approx 0.46 $ MHz for the Na(0) site and $\nu_q \approx 0.31 $ MHz for the Na(1) site, similar to under-doped NaFe$_{0.983}$Co$_{0.017}$As.\cite{Oh.13} The difference in $\nu_q$ for Na(0) and Na(1) is an indication of their different chemical environments.  

We also observed inequivalent Na sites for $x=0.24$ and 0.39 at room temperature, shown in Fig.~\ref{Fig.2}\,(b). For the central transition, a narrow peak with relatively small spectral weight, corresponding to the Na(0) site, is observed in both spectra, while a significantly broadened peak associated with large relative spectral weight is also observed. We conclude that the large, broadened spectral components are contributed by Na sites that have one or more of the four NN sites occupied by Cu dopant(s). We call these sites ``impurity sites'' since they appear as an obvious extension of the components of the spectra for $x=0.13$ and 0.18 but are not well-resolved owing to extensive disorder. For $x=0.39$  there are satellites of the impurity central transition positioned symmetrically with respect to the central transition, providing a clear signature of their quadrupolar origin. Comparison of the wings of $x=0.24$ and 0.39 spectra, {\it i.e.} spectral weight farther from the central transition, suggests that there are satellites from two types of impurity components, one with larger $\nu_q$ associated with  different bonding configurations for Na, as sketched in Fig.~\ref{Fig.1}\,(b).  Correspondingly, we have modeled the full room-temperature $^{23}$Na spectrum of $x=0.39$ with 7 gaussian components, as shown by the black dashed curves in Fig.~\ref{Fig.2}\,(b), where the broad central peak is composed of two gaussians at the same frequency, but different linewidths, and with satellite peaks fitted on each side of the central peaks. In the case of $x =0.24$, only the central transition could be analyzed and so three gaussian components were used to fit the central peaks from Na(0) and two impurity components. For \textit{x} = 0.39, the fact that the linewidths of quadrupolar satellites and central transition only differ by $\sim$50 kHz for the main component of the impurity sites, rules out that their origin is from chemical disorder giving a distribution of EFG's and so unambiguously indicates that the disorder is magnetic. Also from measurements at NHMFL of field dependence of the linewidths with external magnetic field \textit{$H_0$} = 10, 14, and 16 T, we showed that the broadening of the central peaks for $x = 0.13,$ $0.18, $ and $0.24$ compounds scales with magnetic field and consequently confirm their magnetic origin.

We compared the normalized spectral weight of the central transitions in Fig.~\ref{Fig.2} to a binomial model \cite{Nin.10,Mou.13}  for Na(0), Na(1), and Na(2) sites, specified in Fig.~\ref{Fig.1}\,(b).  Assuming a random process, these are given by a probability distribution $P(n,x) = {{4}\choose{n}}x^{n}(1-x)^{4-n}=\frac{4!}{n!(4-n)!}x^{n}(1-x)^{4-n} $ with $n$ being the number of Fe sites substituted for by Cu that are nearest neighbor to a Na, where $x$ is the Cu concentration. As shown in Fig.~\ref{Fig.3}, the normalized spectral weight of Na(0), Na(1), and Na(2) sites for $x=0.13$ and 0.18 fit relatively well to the binomial model. However, for the samples with $x$ = 0.24 and 0.39, we see that the normalized spectral weight of impurity sites far exceeds the predicted value given by the model where impurity is defined as the sum of Na(1) and Na(2).  Additionally, the spectral weight of Na(0) is significantly less than what is predicted by this model.

The good agreement between experimental data and the binomial model for $x=0.13$  and 0.18  indicates that the Cu substitution is dominated by a random process at low doping. However, it cannot be the main mechanism  responsible for Cu distribution in the two samples with higher Cu concentration. It was reported by Song \textit{et al.}~\cite{Son.16}, based on their neutron scattering and high resolution TEM measurements, that Fe and Cu atoms start to form a real space stripe-like ordered structure in NaFe$_{1-x}$Cu$_{x}$As as $x$ approaches 0.5. Therefore it is possible that for $x=0.24$ and 0.39 the Fe and Cu atoms have already begun to form Fe-Cu clusters which continue to grow in size and connect with each other until stripes are finally formed as $x$ reaches 0.5.   Wang \textit{et al.}~\cite{Wan.17} proposed a similar model of Fe-Cu stripe ordering to explain the breaking of four-fold rotational symmetry in Ba(Fe$_{1-x}$Cu$_{x}$)$_2$As$_2$.  We have investigated this formation of AFM clusters using NMR Knight shift and spin-lattice relaxation measurements.

\begin{figure}
	\includegraphics[scale = 0.6]{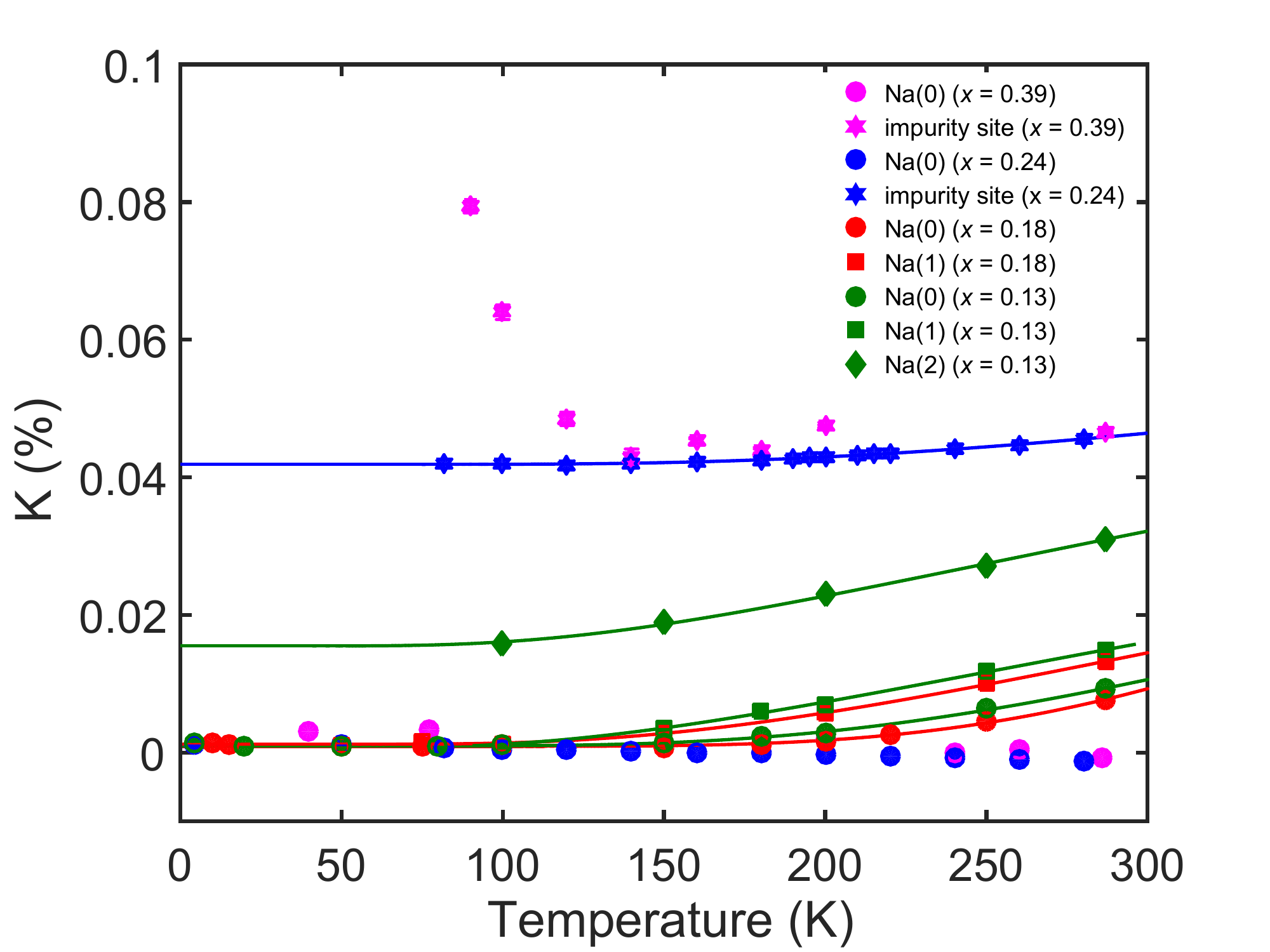}
	\caption{\label{Fig.4} Temperature dependence of the  $^{23}$Na Knight shift, $K(T)$, from different Na sites in $x=0.13$, 018, 0.24, and 0.39 with the external field $H_0 = 14$ T $\parallel$ $c$-axis. Solid curves are fits to an activation process $K_s  \propto\mathrm{exp}(-\Delta_{pg}/k_BT)$ where $\Delta_{pg}$ is a spin pseudogap discussed in the main text. The Knight shift for $x=0.39$ was determined from a single gaussian fit to the central transitions of the Na impurity site, and therefore the corresponding $K(T)$ of the impurity site is only an approximation for these components of the spectrum that substantially increases in width with decreasing temperature.}
\end{figure}

\begin{figure}
	\includegraphics[scale = 0.43]{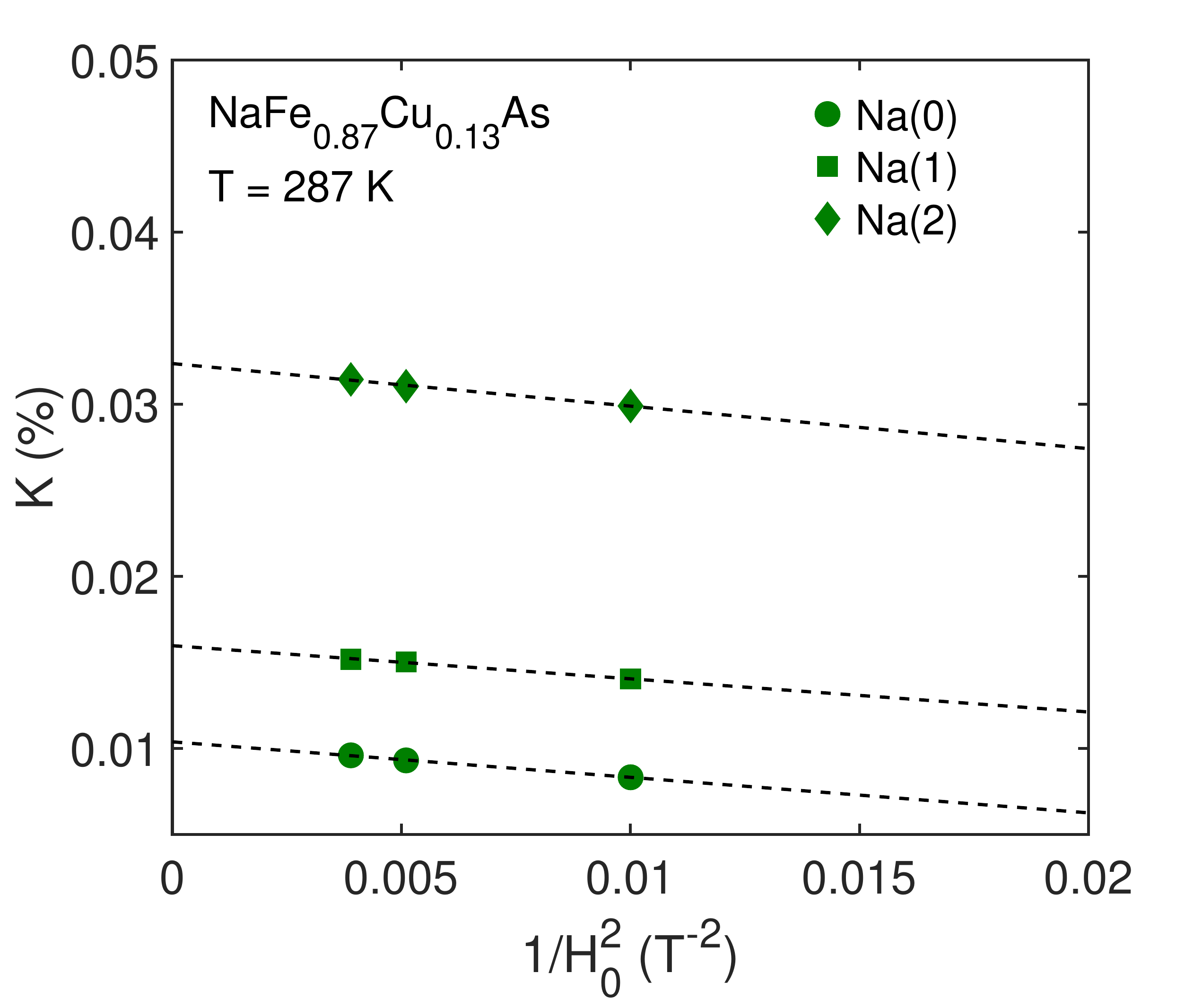}
	\caption{\label{Fig.5} $^{23}$Na frequency shift $K$ plotted as a function of inverse square of magnetic field $1/{H_0}^2$ for central transitions of Na(0), Na(1), and Na(2) sites in the $x = 0.13$ composition. The measurements were done with $H_0 = 10, 14,$ and $16$ T $\parallel$ $c$-axis, and $T$ = 287 K. The extrapolation to $1/{H_0}^2 = 0$ for linear fits (dashed lines) to the data gives the Knight shift. The slopes of the linear fits are close to zero, indicating that the second order quadrupolar contribution to the frequency shift is negligible.}
\end{figure}

\begin{figure}
	\includegraphics[scale = 0.44]{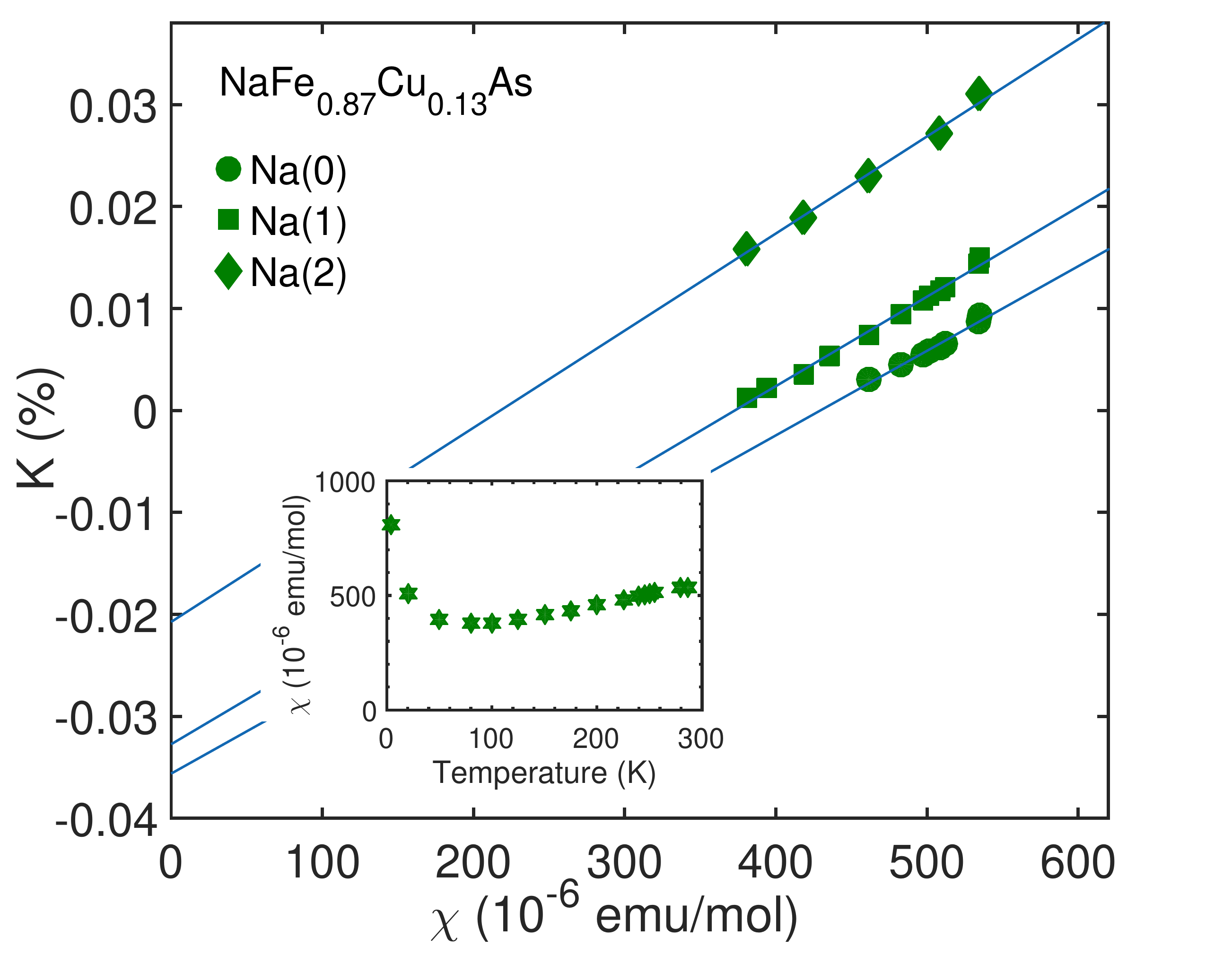}
	\caption{\label{Fig.6} $^{23}$Na Knight shift $K$ plotted as a function of magnetic susceptibility $\chi$ in the range: 200 K $\leq$ $T$ $\leq$ 300 K for the Na(0) site, and  100 K $\leq$ $T$ $\leq$ 300 K for the Na(1) and Na(2) sites for $x=0.13$. The results of a linear fit (solid line) determine the hyperfine coupling constants $A_{hf}$ for Na(0), Na(1), and Na(2) sites. Inset: Temperature dependence of magnetic susceptibility for $x = 0.13$ with the external field $H = 5$ T $\parallel$ $ab$-plane.~\cite{Wan.13}}
\end{figure}

%
 \section{Knight shift}
The temperature dependence of the $^{23}$Na NMR frequency shift for the central transition (1/2 $\leftrightarrow$ -1/2) of the different Na sites is shown for all samples in Fig.~\ref{Fig.4}. The frequency shift is defined by $K(T) = \triangle\omega(T)/\omega_L=(\omega(T)-\omega_L)/\omega_L$, where $\triangle\omega$ is the measured total frequency shift from the Lamor frequency, 
\begin{equation}\label{Eq.1}
K(T) = K_s(T) + K_{orb} + K_{q}.
\end{equation}
\noindent
Here \textit{$K_s$, $K_{orb}$,} and \textit{$K_{q}$} represent the electronic spin, orbital, and quadrupolar contributions to the frequency shift respectively. The sum of $K_{s}$ and $K_{orb}$ is generally referred to as the Knight shift. Both $K_{orb}$ and $K_{q}$ are temperature independent in the absence of a structural transition, and the latter is expected to be negligible for the central transitions of Na(0) and Na(1) sites. This follows from the symmetry of the crystal structure and the configurations of neighboring atoms near Na for which the asymmetry parameter can be expected to be small, at least for those two sites. However, to investigate this quantitatively we rely on the fact that $K_q$ for the central transition, if it exists, is proportional to the inverse square of magnetic field $1/{H_0}^2$ while the Knight shift is magnetic field independent. Focusing on the $x = 0.13$ composition, which has the best resolution, we plot the frequency shift $K(T)$ versus $1/{H_0}^2$ for $H_0 = 10, 14,$ and $16$ T in Fig.~\ref{Fig.5}. The dashed lines are linear fits to the data and the extrapolation to  $1/{H_0}^2 = 0$ gives the Knight shift. As shown in Fig.~\ref{Fig.5}, the slopes of the linear fits are similar, and very small for all sites, confirming that the effect of second order quadrupolar shifts can be neglected.

The total electronic spin contribution $K_s(T)$ can have temperature dependence related to the hyperfine coupling of the nucleus to either conduction or localized electron spins.  In our composition and temperature range the electrical resistivity increases with decreasing temperature by one ($x=0.18$) to two ($x=0.39$) orders of magnitude which is not evident in the temperature dependence of $K_s$.  Consequently conduction electron spin contributions to $K_s(T)$ are negligible. In the present case the temperature dependence is from localized spins associated with a spin pseudogap which we will discuss later, after analysis of the hyperfine field that couples these electronic spins to $^{23}$Na. 

We can extract the hyperfine field by comparing Knight shift with the measured total magnetic susceptibility, $\chi(T) = \chi_s(T) + \chi_{0}$, separating the temperature dependent (spin) and temperature independent contributions. Focusing on the most dilute doped crystal, $x=0.13$, we  plot $K(T)$ versus the magnetic  susceptibility  $\chi(T)$ for the Na(0), Na(1), and Na(2) sites in Fig.~\ref{Fig.6}, where temperature is an implicit variable. The slope of such a plot gives the hyperfine field at those Na sites that are coupled to the temperature dependent susceptibility, $\chi_s(T)$, shown in the inset of Fig.~\ref{Fig.6}. The linear behavior in the main figure is ascribed to $K_s(T) = A_{hf}\chi_s(T)/N_A\mu_B$, where $\chi_s$ is the uniform spin susceptibility, $A_{hf}$ is the hyperfine coupling constant, $N_A$ is Avogadro's number, and $\mu_B$ is the Bohr magneton.  We find that $A_{hf} = 4.7 \pm 0.7$ kOe/$\mu_B$, $4.9 \pm 0.3$ kOe/$\mu_B,$ and $5.3\pm0.3 $ kOe/$\mu_B$ for Na(0), Na(1), and Na(2) sites respectively, suggesting that the hyperfine coupling is the same for all sites. This could indicate that at the Cu site, there might exist a non-zero magnetic moment that couples to Na through hyperfine coupling for the $x = 0.13$ composition, as expected when Cu is in the Cu$^{2+}$ state at low doping. The figure inset also indicates a substantial upturn in susceptibility at low temperatures ($T < 50$\,K)  for the $x=0.13$ composition that is not matched by a corresponding increase in $K(T)$. This Curie-Weiss  upturn is often associated with local moments induced by defects or impurities, and has been reported in many Fe-based compounds.\cite{Wan.12,Xia.12,Cao.09} Since the susceptibility upturn has no obvious hyperfine coupling to the $^{23}$Na nucleus manifested in either NMR shift or linewidth, it is likely extrinsic and can be ignored.

The  orbital shifts for all compositions at Na(0) are negative but very small in magnitude, $|K_{orb}| \leq 0.002\%$.  However, for Na(1), Na(2), and impurity sites, $K_{orb}$ increases positively and systematically with increased doping, Fig.~\ref{Fig.4} and Fig.~\ref{Fig.6}.
There are two principal sources for orbital shifts, often referred to as chemical shifts.  Small negative shifts come from diamagnetic screening currents; positive shifts are commonly associated with nearly degenerate atomic level splittings that are sensitive to the local chemical environment. The latter contributions to the Van Vleck susceptibility are expected in second order perturbation theory and they correspond to NMR frequency shifts that are necessarily positive. We identify the significantly larger increase in $K_{orb}$ for $x=0.24$ and 0.39 at sites having NN Cu dopant atoms with the valence change proposed by Song \textit{et al.} of Cu$^{2+}$ to Cu$^{1+}$, taking one electron from its neighboring Fe ion, and consequently changing the Fe ion from its original Fe$^{2+}$ to Fe$^{3+}$ state.~\cite{Son.16,Yu.13}  Additionally at low temperatures below $\approx80$\,K for the impurity site at the highest doping of $x=0.39$, the frequency shift has a substantial increase that cannot be understood as a contribution to $K_{orb}$. We identify this temperature dependence with contributions to $K_s$ by magnetic clusters, which are precursor to the AFM discussed by Song \textit{et al.}~\cite{Son.16}  Consistently, as we noted in section III, there is a substantial increase in magnetic contributions to the spectral linewidth of $x=0.24$ and 0.39, also an indication of the formation of AFM clusters. Our interpretation agrees with  measurements of a short nm-scale correlation length from neutron scattering.~\cite{Son.16} However, with regard to Fig.~\ref{Fig.1}, we clarify that  there is no NMR evidence for a phase transition to static magnetic order for $x\leq 0.39$

\begin{figure}
	\includegraphics[scale = 0.5]{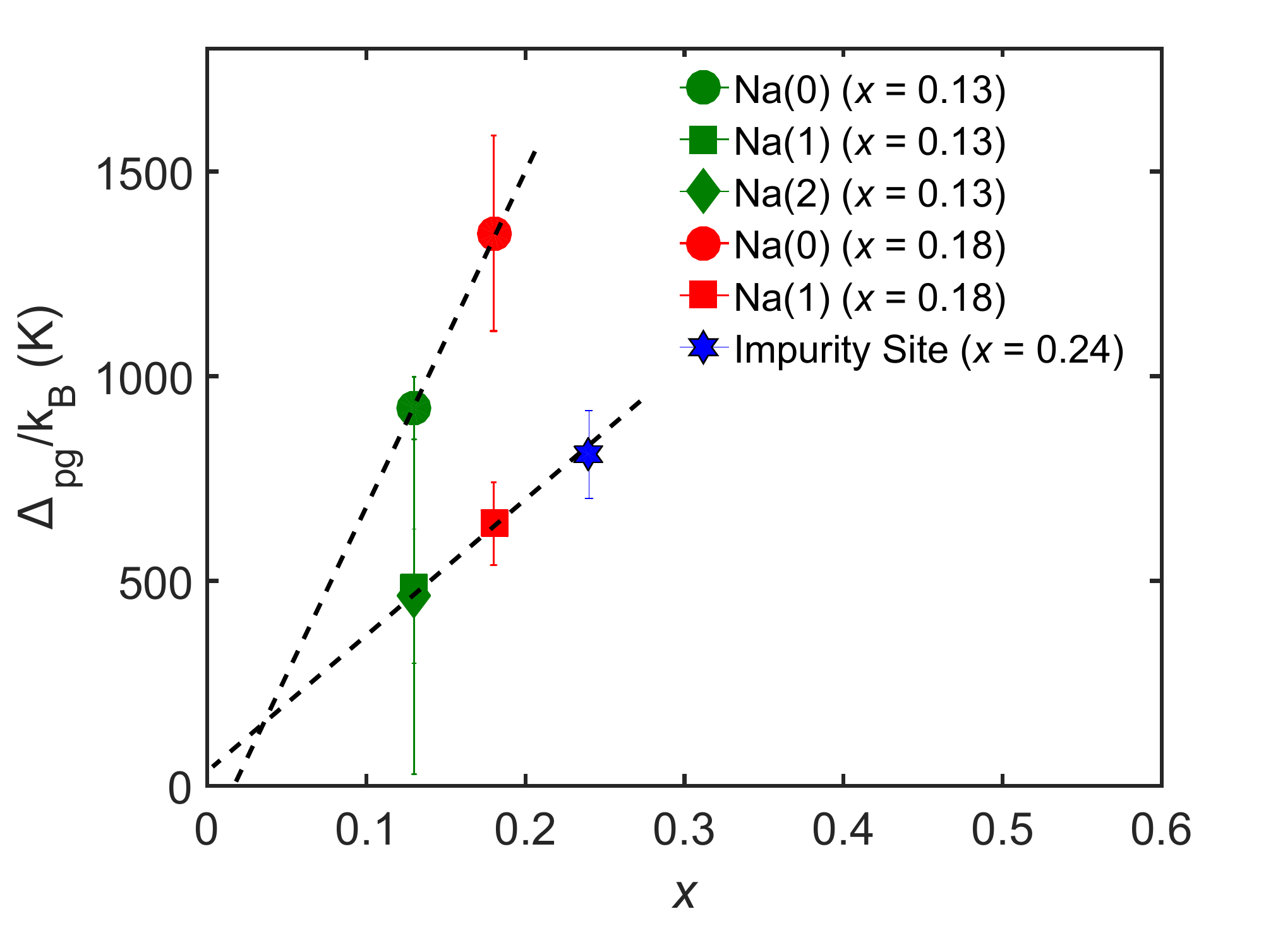}
	\caption{\label{Fig.7} Dependence of $\Delta_{pg}/k_B$ on Cu doping suggestively decreasing toward zero for the parent compound. The dashed lines are guide for the eye only.} 
\end{figure}

Returning to the temperature  dependence of $K_s(T)$ at high temperature, we find this to be consistent with an activated process for all Na sites, for the compositions $x=0.13$ and 0.18, and for the impurity site in $x=0.24$ and 0.39, corresponding to a temperature-dependent suppression of $\chi_{s}(T)$ with decreasing temperature. The reduction of $\chi_{s}(T)$ and $K(T)$ is caused by a spin pseudogap $\Delta_{pg}$, and is common among various iron-pnictide compounds.~\cite{Nin.08,Ahi.08,Oh.11} The solid lines in Fig.\ref{Fig.4} are the fits of the data to the activation form, $K(T) = A\times\mathrm{exp}(-\Delta_{pg}/k_BT)+B$, where $k_{B}$ is the Boltzmann constant, and $A$ and $B$ are constants. The values of $\Delta_{pg}/k_B$ obtained by fitting the data are plotted in Fig.~\ref{Fig.7} versus Cu concentration for Na(0) and Na(1) sites. The magnitude of $\Delta_{pg}$ appears to increase proportional to Cu doping suggesting a correlation between the magnitude of the pseudogap and the increased magnetic neutron scattering intensity.\cite{Son.16} For Na(0) sites in $x=0.24$ and 0.39, the pseudogap is sufficiently large that it does not have an observable effect on $K(T)$ in our accessible range of temperature.

\begin{figure}
	\includegraphics[scale = 0.45]{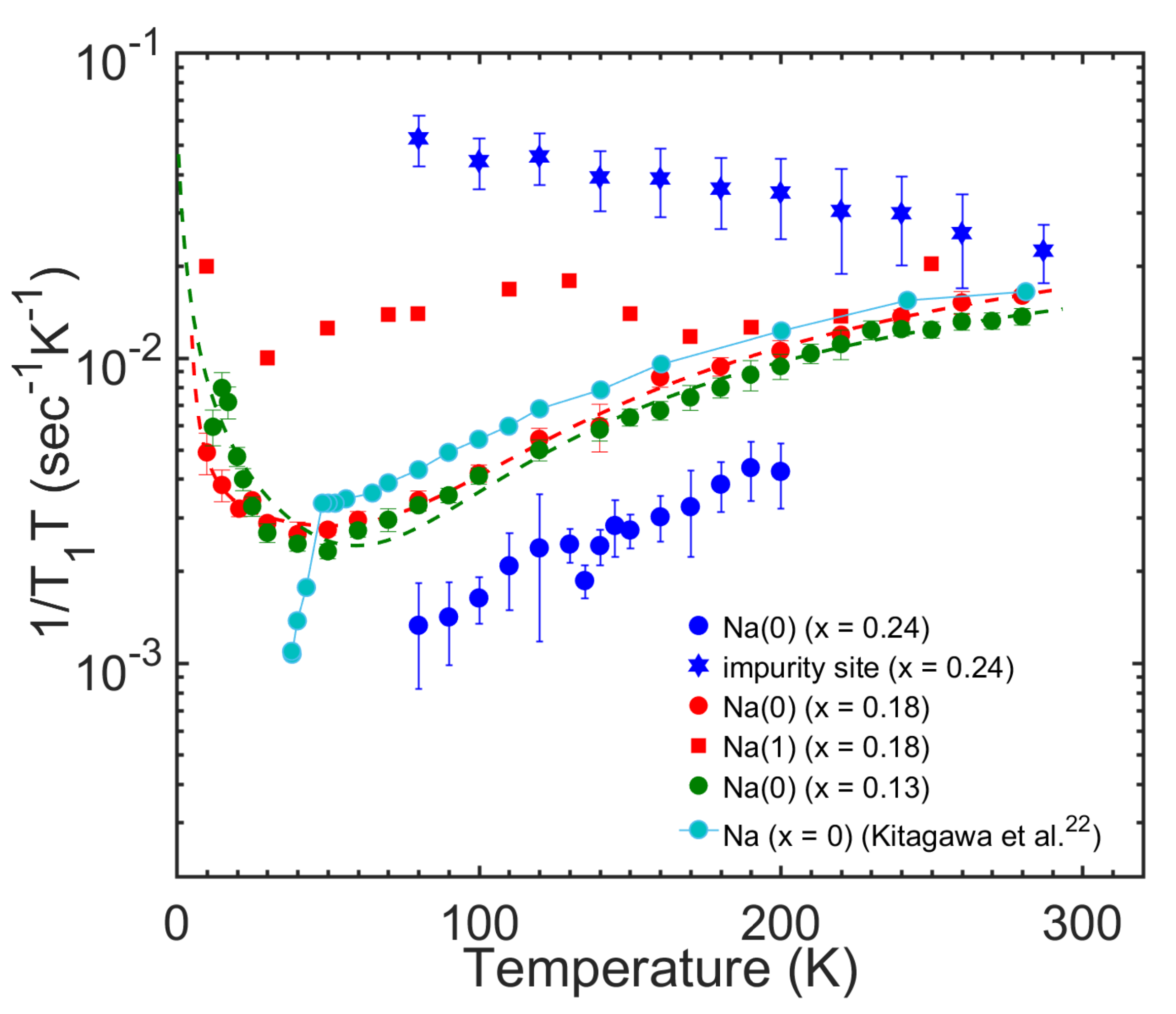}
	\caption{\label{Fig.8} The spin-lattice relaxation rate $1/T_1T$ measured for the central transitions of the Na(0) site in samples with $x$ = 0.13, 0.18, and 0.24; the Na(1) site in the sample with $x$ = 0.18;  and the impurity site in the sample with $x$ = 0.24, at $H_0 = 14$\,T $\parallel$ $c$-axis, and for the parent compound NaFeAs, included for comparison.~\cite{Kit.11} The dashed lines represent two fits to a combination of pseudogap $(1/T_1T)_{pg}$ and Curie-Weiss $(1/T_1T)_{cw}$ terms,~\cite{Nin.10} $1/T_1T = (1/T_1T)_{pg}+(1/T_1T)_{cw}
			= (\alpha + \beta exp(-\Delta/k_BT)) + C/(T+\theta)$, where $\Delta$ is the energy gap. In the main text this is compared with $\Delta_{pg}$ obtained from the Knight shift data. The solid line is guide for the eye only.}
\end{figure}

\begin{figure}
	\includegraphics[scale = 0.44]{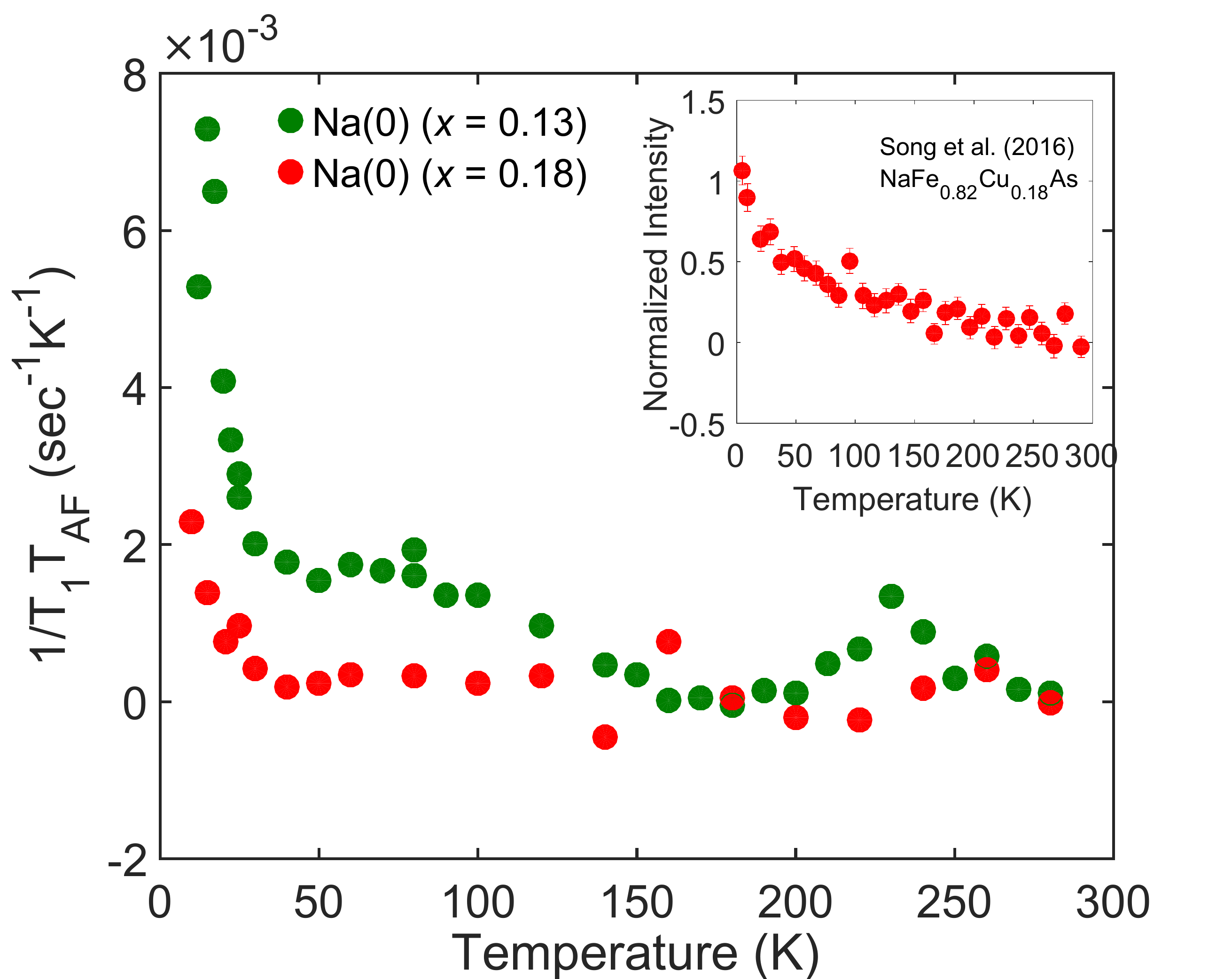}
	\caption{\label{Fig.9} Temperature dependence of $(1/T_1T)_{cw}$ for $x=0.13$ and 0.18. $(1/T_1T)_{cw}$ is obtained by subtracting the pseudogap term $(1/T_1T)_{pg}$ from $1/T_1T$. Inset: Temperature dependence of normalized intensity of the neutron scattering at \textbf{q}$_{AF}$ = (1,0,0.5) for $x=0.18$ from Song {\it et al.}~\cite{Son.16}}
\end{figure}
\section{Spin-lattice Relaxation}

Additional information on spin dynamics of the system can be obtained from the temperature and doping dependence of $1/T_1T$, which is related to the wave vector integral of the low frequency component of spin fluctuations given by the imaginary part of the dynamic susceptibility at the Larmor frequency, $\chi^{''}(\textbf{q},\omega_L)$, 
\begin{equation}\label{Eq.2}
	1/T_1T \sim \sum\limits_{\textbf{q}}|\gamma_nA(\textbf{q})|^2\frac{\chi^{''}(\textbf{q},\omega_L)}{\omega_L}. 
\end{equation}
Spin-lattice relaxation is closely related to  neutron  inelastic scattering intensity extrapolated to very low energy transfer. The form factor $A($\textbf{q}$)$ of the hyperfine interaction between the electronic and nuclear spins at $q =0$ is proportional to the spin part of the Knight shift $K_s$. In Fig.~\ref{Fig.8}, we show $1/T_1T$ of $^{23}$Na spectral components of the central transitions for the crystals $x=0.13$, 0.18, and 0.24. For comparison, we also include the $1/T_1T$ results of $^{23}$Na for the parent NaFeAs compound from Kitagawa \textit{et al.}~\cite{Kit.11} The results at the Na(1) site in $x=0.18$ and the impurity site in $x=0.24$ have a clear enhancement of the imaginary part of dynamic susceptibility $\chi^{''}(\textbf{q},\omega_L)$. The temperature dependence of $1/T_1T$ of Na(0) sites in both $x=0.13$ and 0.18 shows pseudogap behavior above 50\,K, and Curie-Weiss behavior below 50\,K due to antiferromagnetic spin fluctuations that we will further discuss later in this section. Solid curves are the best fits with a combination of pseudogap $(1/T_1T)_{pg}$ and Curie-Weiss $(1/T_1T)_{cw}$ terms,~\cite{Nin.10} 

\begin{equation}
\label{Eq.3}
1/T_1T = (1/T_1T)_{pg}+(1/T_1T)_{cw}
= (\alpha\times exp(-\Delta/k_BT)+\beta) + C/(T+\theta).
\end{equation}
\noindent The pseudogap term has the same activation form we fit the frequency shift data to that we have discussed earlier. 
\begin{figure}
	\includegraphics[scale = 0.42]{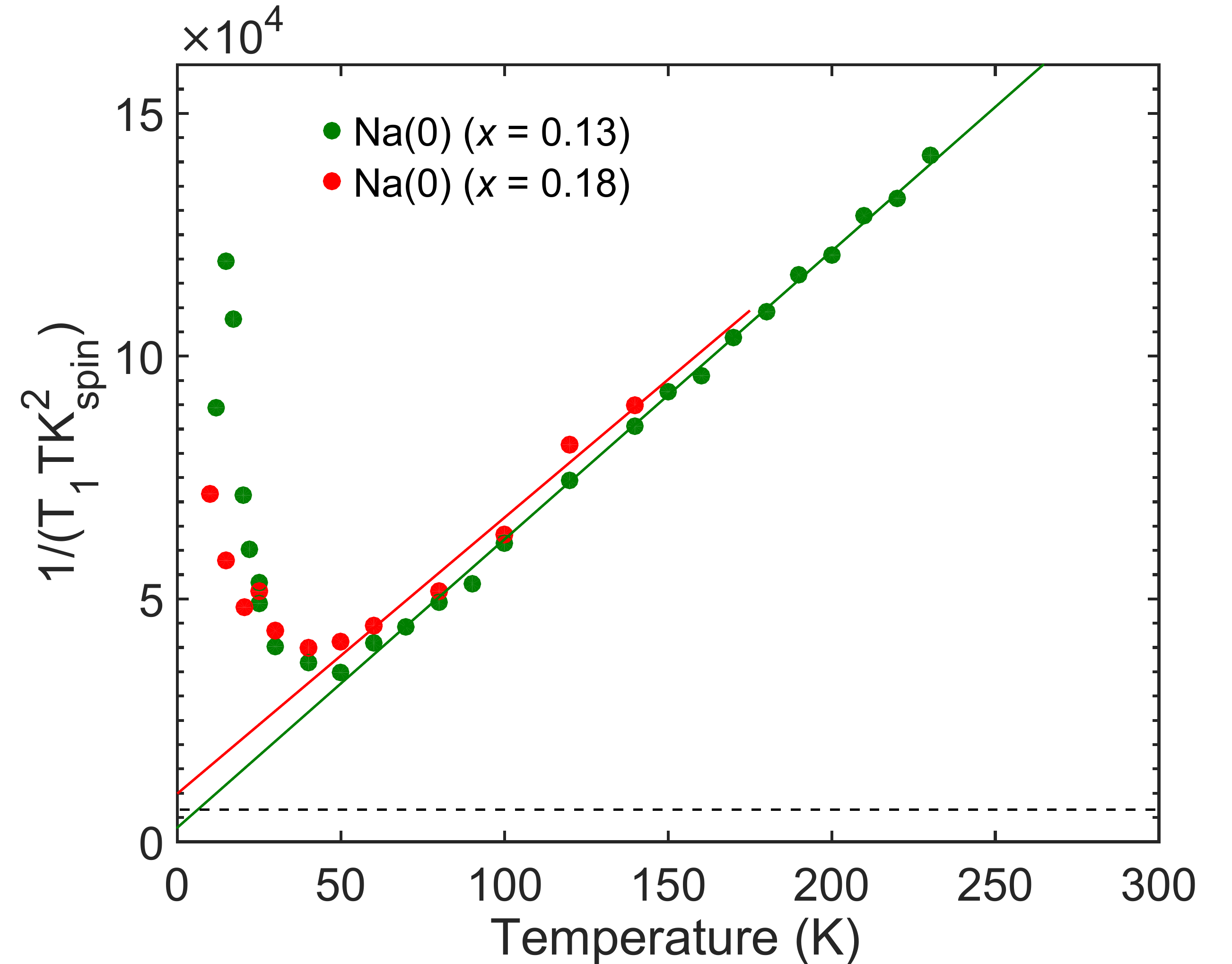}
	\caption{\label{Fig.10} Temperature dependence of $1/T_1TK_s^2$ for $x=0.13$ and 0.18. The solid lines are best linear fits over the temperature range above 50\,K. For comparison we include the dashed line corresponding to the Korringa constant calculated for $^{23}$Na.}
\end{figure}

We first discuss our results in the high temperature regime, $T>50$\,K.  The relaxation rate at the Na(0) sites have an activated temperature dependence which we identify with the spin pseudogap, qualitatively similar to the behavior we observe for $K_s(T)$ including a decrease in magnitude passing from $x=0.18$ to 0.24.  This suppression of $1/T_1T$, and correspondingly $K_s$, at the Na(0) site is due to depletion of electronic states consistent with the observed increase in the resistivity reported by Song \textit{et al.}~\cite{Son.16} in this composition range. On the other hand, for the Na sites with Cu near neighbors there is a marked increase in the relaxation rate in this higher temperature region  that we associate with  growth of magnetic clusters which we have also inferred from the NMR linewidth, Fig.~\ref{Fig.2}, and neutron scattering results.~\cite{Son.16}  Our fit to an activation form of the relaxation rate at the Na(0) sites gives an energy gap of 283\,K and 344\,K for $x=0.13$ and 0.18 respectively,  significantly smaller than $\Delta_{pg}$ determined from fitting $K_s(T)$ with the same activation formula for the samples shown in Fig.~\ref{Fig.7}. This difference can be attributed to the wave vector \textbf{q} and frequency dependence of the pseudogap, however, the origin of the pseudogap behavior shown in both Knight shift and $1/T_1T$ is not yet established.

At low temperatures $T$ $<$ 50\,K, the Na(0) sites for $x=0.13$ and 0.18  have an upturn in $1/T_1T$ which is not evident in $K_s$.  These spin-lattice relaxation measurements provide a sensitive measure of spin dynamics at the Larmor frequency and at non-zero wave vector (Eq.~\ref{Eq.2}), that are not evident in the Knight shift providing an  indication of the early stages of magnetic cluster formation which does not contribute to the static uniform susceptibility $\chi(\textbf{q} = 0)$ probed by $K_s$. We obtain the temperature dependence of the Curie-Weiss term $(1/T_1T)_{cw}$ of $x=0.13$ and 0.18 by subtracting the pseudogap term $(1/T_1T)_{pg}$ from $1/T_1T$ and plot their temperature dependences,~Fig.~\ref{Fig.8}. The results are qualitatively consistent with the behavior of the magnetic order parameter from elastic neutron scattering measurements for $x=0.18$ at \textbf{q}$_{AF}$ = (1,0,0.5),~\cite{Son.16} shown in the inset of the figure, suggesting that the contribution at \textbf{q}$_{AF}$ = (1,0,0.5) is most likely responsible for $(1/T_1T)_{cw}$. We conclude that antiferromagnetic spin fluctuations are the origin of the Curie-Weiss behavior in the $1/T_1T$ results for $T$ $<$ 50\,K.

The Knight shift is proportional to the hyperfine field  and the relaxation rate depends on the square of the hyperfine field.  Consequently, one might expect that that a common origin for contributions to $\chi$ and $\chi''$ would be revealed by plotting $1/(T_1TK_s{^2})$ as a function of temperature. One example of such behavior is the case of a simple metal where this combination is called the Korringa constant and for $^{23}$Na this corresponds to the dashed horizontal line in Fig.~\ref{Fig.10}. It is interesting that the results $1/(T_1TK_s{^2})$ for Na(0) in the two compositions $x=0.13$ and 0.18,  shown at high temperatures in Fig.~\ref{Fig.10}, have a simple proportionality with temperature  in the region attributable to the pseudogap, with an intercept at $T=0$ that is small and of the order of the Korringa constant.  Despite the rather small, but non-zero conductivity of NaFe$_{1-x}$Cu$_{x}$As at low doping, it is worthwhile noting that in the literature there are studies of AFM and FM electron correlations  using the Korringa ratio, $\alpha = \frac{S_0}{K_s^2T_1T}$, where $S_0$ is the Korringa constant.~\cite{Mor.63,Cui.16} For this case, one can in principle determine whether the electron correlation is AFM or FM (AFM for $\alpha>$ 1; FM for $\alpha<$ 1). However, applying this type of analysis to our system is problematic particularly as x approaches 0.5, where our compound becomes non-metallic.~\cite{Son.16}

\section{conclusion}
In summary, we have reported the first NMR investigation of heavily doped single crystals of NaFe$_{1-x}$Cu$_{x}$As with $x=0.13$, 0.18, 0.24, 0.39. Our $^{23}$Na NMR spectra reveal the existence of inequivalent $^{23}$Na sites corresponding to the distribution of Cu dopants having zero, one, or two nearest neighbors substituting for Fe in the Fe sublattice.  For the higher Cu concentration samples, $x$ = 0.24 and 0.39, the spatial distribution deviates from random  consistent with the observation of real space Fe-Cu stripe-like-ordered structure shown by neutron scattering and high resolution TEM measurements.~\cite{Son.16,Wan.17} Our measurements of the quadrupolar frequency, $\nu_Q$, and orbital Knight shift, $K_{orb}$, are sensitive to chemical bonding of Na to near neighbors and varies with composition, $x$,  changing substantially between $x = 0.18$ and 0.24, exactly the range for increased magnetic disorder evident in the NMR linewidth which we  associate with magnetic cluster formation.  Our temperature dependent measurements of the NMR spin-dependent Knight shift, $K_s$, reveal pseudogap behavior at various of the Na sites, and we find that increasing Cu doping selectively removes the effect of the pseudogap at certain sites.  We interpret these changes as a consequence of a valence change from Cu$^{2+}$ to Cu$^{1+}$, and concomitantly Fe$^{2+}$ to Fe$^{3+}$, as $x$ increases, as suggested by both theoretical calculation and other experimental results.~\cite{Yu.13,Son.16} The spin-lattice relaxation rate $1/T_1T$ measurements,  combined with the Knight shift $K_s$,  show AFM electron correlations to be the source of AFM fluctuations in the temperature range $T < 50$\,K, likely at the wave vector, ~{\bf q}$_{AF}$ = (1,0,0.5), identified by neutron scattering,~\cite{Son.16}.  Most importantly we report evidence for the systematic development of AFM clusters with increasing Cu dopant,  precursory to the reported AFM striped state.~\cite{Son.16} But in the composition range $x \leq 0.39$ there is no long range static magnetic order.

\section{acknowledgment}
We thank Arneil Reyes for use of facilities at NHMFL, and  Weiyi Wang and Chongde Cao for their contributions to the crystal growth and characterization. Research was supported by the U.S. Department of Energy (DOE), Office of Basic Energy Sciences (BES), Division of Material Sciences and Engineering under Award No. DE-FG02-05ER46248 (WPH) and DE-FG02-05ER46202 (PD), and the NHMFL by NSF and the State of Florida. The single crystal growth efforts at Rice were supported by the U.S. DOE, BES under Grant No. DE-SC0012311.  Part of the materials work at Rice was supported by the Robert A. Welch Foundation under Grant No. C-1839.

\bibliography{maintext}

\end{document}